# Experimental study of the spatio-temporal development of meter-scale negative discharge in air

P O Kochkin[1], A P J van Deursen[1], U Ebert[1,2]

[1] Department of Electrical Engineering, Eindhoven University of Technology, POBox. 513, NL-5600 MB Eindhoven, The Netherlands

[2] Department of Applied Physics, Eindhoven University of Technology, and Centre for Mathematics and Computer Science (CWI), POBox 94079, NL-1090 GB Amsterdam, The Netherlands

E-mail: p.kochkin@tue.nl

**Abstract.**

We study the development of a negative discharge driven by a Marx generator of about 1 MV in an air gap of 1 up to 1.5 meter, at standard temperature and pressure. We show the evolution of the discharge with nanosecond-fast photography together with the electrical characteristics. The negative discharge develops through four well-distinguished streamer bursts. The streamers have different velocities and life times in different bursts. The last burst triggers a positive inception cloud on the positive grounded electrode and a burst of positive counter-streamers emerges. The pre-discharge then bridges the gap and leaders grow from both electrodes. Finally a spark is formed. Looking closer into the pre-ionized zone near the cathode, we find isolated dots which are potential branching points. These dots act as starting points for positive streamers that move towards the high-voltage electrode. We also find such phenomena as space leaders and leader stepping in our laboratory sparks.

**1. Introduction**

Natural lightning is a very complex phenomenon evolving on multiple temporal and spatial scales. The stepping of lightning leaders, space stems, dart leaders as well as the energetic radiation from thunderstorms such as Terrestrial Gamma-Ray Flashes, electron beams and electron-positron beams are far from being fully understood. Furthermore a wealth of discharge phenomena was discovered above thunderclouds, called Transient Luminous Events (TLE's), including sprite discharges. All these phenomena are currently subject of international research, geophysical conferences, workshops and summer schools. However, natural lightning events are erratic and difficult to access for precise observations.

In previous work [1] we have presented images and analysis of metre length positive discharges, and of their energetic radiation. The same source was a standard lightning impulse generator of 2 MV maximum. Here we focus on the structure and evolution of negative discharges, having in mind that the majority of cloud-to-ground lighting flashes have negative polarity. We find stepping phenomena, space stems and energetic radiation as well, but also discharge features that to our knowledge have not yet been found in lightning. Therefore we believe that our study is of interest for lightning research as well as for lightning protection and high-voltage technology.

Only few studies [2,3] have been performed recently on metre-scale laboratory discharges in STP air because these require a large experimental facility, modern fast cameras and careful electromagnetic shielding of measuring equipment. A recent overview is given in [4]. The current understanding of long discharges is still mostly based on streak camera observations [5] performed in 1980's. Reess et al. [6] performed an experimental study of negative discharges in a meter-scale gap in 1995 with streak photography. Our voltage amplitude of 1.2 MV and gap length of about 1 m are of the same order. The high-voltage pulse shape differ: 1.2/50 μs versus 0.3/2000 μs in [6]. We use two conical electrodes, instead of a point-plane gap. The conical ground electrode defines a clear starting point for the counter-discharge and allows the current through that electrode to be measured. Two high-resolution, nano-second fast intensified CCD cameras have been used. Their adjustable shutter delay and opening times provide the opportunity to take two subsequent images, which then show the discharge growth, its speed and direction. The synchronised measurement of the electrical parameters gives the stages of the discharge: formation of the inception cloud, streamers bursts, streamer encounters and even leader formation and stepping. With the aid of fast computer storage of data and images, more than 3000 discharges are available for analysis. This large number allows to discern particular and general phenomena.

In this paper we show high-quality images of negative discharges in STP air, combined with the electrical characterization of the discharges. In our setup the streamers emerge in bursts as the voltage rises. We follow the negative discharge development from the first light emission up to the final breakdown, first (Sect. 3.1) in time-integrated images as in [1] or streak photography [6]. The time sequences with two cameras provide more information (Sect. 3.2). We present detailed images of the chaotic phenomena at the high voltage electrode over the fourth streamer burst (Sect. 3.3). Observations on the leader phase (Sect. 3.4) and a brief comparison with phenomena observed in nature (Sect. 3.5) follow.

**2. Experimental setup**

The setup is similar to the one described in [1]; for the sake of completeness it is represented again in Figure 1. The 2 MV Marx generator delivers a high-voltage (HV) standard lightning pulse with 1.2/50 μs rise/fall time. The upper voltage limit employed was about 1.2 MV. The generator voltage was measured by a capacitive high-voltage divider. The electrodes of the spark gap are cones. The tip distance was varied between 1 m and 1.75 m. Two Pearson 7427 current probes determined the currents through the HV electrode (cathode) and the grounded electrode (anode). An optical transmission system inside the HV electrode transported the HV current signal. Suitable attenuators and two antiparallel high-speed diodes protect the input of the transmitter. The diodes limit the linear response to 250 A; above this value the transfer is approximately logarithmic. An RG214 cable connected the current probe for the grounded electrode directly to the measuring system. An aluminum disk mounted near each probe minimised the risk of a direct hit by the full 4 kA spark current.

Two Picos4 Stanford Optical cameras were placed side by side at 4 m distance from the spark gap. The cameras contain charge coupled image sensors preceded by a fast switched image intensifier (ICCD). The image intensifier is a micro-channel plate that allows us to adjust the camera sensitivity by varying the applied voltage. The CCD is read out with 12 bit resolution. The camera optical axis was most often directed towards the spark gap centre. In comparison with our previous study on positive discharges [1], optical shielding of the spark gaps in the Marx generator significantly reduced the laboratory background visibility in the images. Appropriate electromagnetic shielding protected the cameras and their communication cables against electro-magnetic interference. The image intensifier amplification has been set to accommodate the light level and range of a particular experiment. Both cameras were adjusted to the same sensitivity, but no absolute calibration has been carried out. Lenses were either Nikon 35 mm F2.8 fixed focus or Sigma 70-300 mm F/4-5.6 zoom. With the telelenses, structures of only half a millimetre in diameter were well resolved near the electrodes. The full optical spectrum was used as emitted by the discharge, as transmitted by the lenses and captured by the ICCD cathode. As mentioned in the Introduction, the cameras were triggered simultaneously and opened their shutters after an adjustable delay, which allowed capturing two subsequent images during one discharge. To smooth the contrast of some images in print, we sometimes applied a gamma correction. The measured light intensity $I_m$ is then scaled between 0 (black) and 1 (white) and is displayed as $I_m^\gamma$ intensity in the figures. A value of $\gamma$ less than one increases the relative intensity of darker pixels. The figure captions mention such a correction when applied. On the cameras leaders appear at least four times brighter than the steamers head traces. The steamer-leader transition is gradual as leaders form out of streamer channels by Ohmic heating of the gas. In this sense, we introduce the term 'pre-leader' for stationary channels that are definitely brighter than the streamer head, but not yet at full leader intensity. But even with the high camera speed, stationary dim phenomena can look bright. Additionally the measured intensity depends on the orientation and direction of movement with respect to the camera axis.

The electrical signal acquisition system consisted of two LeCroy oscilloscopes with 1 GHz bandwidth. The negative edge of the signal from the HV divider triggered the oscilloscopes. One oscilloscope then also triggered both cameras. The differences in the delays caused by the instruments and cables were corrected for to within 1 nanosecond accuracy.

Some remarks on the measured voltage follow: ideally the gap voltage is measured as line integral of the electric field over a straight path between the electrodes. But any object there would interfere with the discharge formation. We use the voltage obtained from the 9 m tall high voltage divider instead, which differs from the gap voltage because of the inductance $L$ in the large loop formed by the divider, wiring and gap. From the measured resonance frequency of 1.7 MHz at gap breakdown and the known 600 pF HV divider capacitance we obtain $L = 15$ μH. Values for the current derivative $dI/dt$ at the high voltage electrode of the order of 1 A/ns will be shown below. To this adds the current contribution from discharges at the wiring towards the electrode. The inductive voltage difference between divider and gap is thus 15 kV or more. Faster current variations also occurred occasionally at the electrode, and then the 20 pF capacitance of the high voltage electrode and nearby wiring act as supply. For these reasons we preferred to use the electrode currents as critical parameter in the analysis. The rather high impedance of the surge generator as current source and the microsecond variation of the voltage made the various stages in the discharge formation discernable.

## 3. Results

We studied the different phases of negative discharge development in detail. First one has the inception cloud, a faint ionized region that develops around the sharp edges and tips of the electrodes. Because of the low light level compared to later phases, we set the camera sensitivity to a high level and zoomed in on the electrodes. With rising voltage and electric field at the cloud surface, streamers emerge from the cloud. In Section 3.1 we show the time integrated images taken with increasing exposure times on a single camera. These images can be compared with those presented earlier for positive discharges in [1]. Short exposure images taken with two cameras in sequence are presented in Section 3.2. When the streamers bridge the gap, those between the electrodes become bright because of increased current and Ohmic heating while streamers in other directions extinguish. The fine structure of the later phases near the HV electrode is discussed in Sect. 3.3. Section 3.4 presents the formation of leaders in our metre-sized gap, which are attached to the electrodes. For gaps of about one metre or less, the final stage is a spark between the electrodes. The spark properties are outside the scope of this paper. The spark does not occur at 1.75 m electrode distance within the time allowed by the surge generator fall time of 50 µs. In Section 3.5 we show the striking similarity in appearance between some discharges in our set-up and tens of kilometer sized sprite discharges [7,8] at high altitude and strongly varying low gas pressure.

*3.1 Sequence of time-integrated images*

Figure 2 presents the negative discharge development in a series of time-integrated images, taken with a single camera, where the camera shutter always opens at the same moment in the voltage waveform, but the exposure time increases from panel to panel. The high-voltage (top) and grounded electrode (bottom) are indicated in the images. The distance between electrodes is 1.27 m between the tips. Every image shows a different discharge under the same conditions, but the images of the discharge development show a large similarity from discharge to discharge. The electrical signals of the voltage (U), cathode ($I_{HV}$) and anode ($I_{GND}$) current are represented in the bottom plot (p); these are the average over 65 discharges. The averaging is allowed because also the electrical characteristics between the individual discharges are very similar. For instance the details of the $I_{HV}$ curve remain recognizable such as the first leap in between 0.2 – 0.3 µs. The trigger for the oscilloscopes and cameras is derived from the voltage as measured by the HV divider (Figure 1). Even for the strongly fluctuating current $I_{HV}$ on the HV electrode, the jitter over different discharges is only 50 ns. On the time axis t = 0 µs corresponds to the start of the voltage pulse. The solid vertical line *z* at t = 0.48 µs in plot (p) indicates the shutter opening time of the camera. Dotted vertical lines (*a*)-(*o*) label the shutter closing times for the corresponding images. Thus, a single image is time-integrated from the starting time *z* until the shutter closing time between 25 (*a*) and 1650 (*o*) ns later. A faint speckle trace connecting the electrodes is visible in the first six images. It is caused by the huge brightness of the spark that appears after the camera shutter is closed. The electronic shutter is good but not perfect; some light of the arc leaks through the image intensifier even when it is off. But the speckle trace shows the final breakdown path even on pre-breakdown images. During the entire sequence we keep the camera sensitivity constant. The linear color coding scheme to the ICCD output is indicated by the color bar on the right. No gamma correction has been applied. The negative inception cloud in not included in this sequence; these images are shown later in Section 3.1.1.

At the beginning, a current of up to 40 A is measured on the high-voltage electrode between the times 0.15 and 0.32 µs, while the voltage rises to 180 kV; see Figure 2 (p). The current leap corresponds to the first streamer burst or corona formation, where the current is the sum for all streamers from the tip and the edges of the protection disk. The high-voltage current decreases to zero at t = 0.35 µs, in spite of the voltage rising to 200 kV. The current zero indicates that the streamer growth stops momentarily. The second cathode current leap with the maximum amplitude 75 A and its associated streamer burst appears between t = 0.35 to 0.47 µs. The streamers are slightly visible in image (*a*) at 35 cm distance from the high-voltage electrode tip. Thus, two small streamer bursts with corresponding current pulses on the HV electrode preceded the sequence shown in Figure 2.

In Figure 2 (*a*) – (*c*) we see the third streamer burst. Because of the time-integration, image (*b*) comprises all light of (*a*) and so on. While the applied voltage keeps increasing, at t = 0.7 µs the fourth and largest streamer burst starts. The corresponding image (*d*) shows a well-developed negative corona near the high-voltage electrode where the air is already ionized by previous streamers. From the continuity of the traces, we conclude that the outer-most streamer tracks are an extension of those existing earlier, in agreement with [9]. Some of these streamers propagate quite far across the gap until close to the grounded electrode (images (*d*) – (*g*)). In image (*e*) a faint positive inception cloud [10–12] is visible near the grounded electrode, because the approaching negative streamer charge raises the electric field at this electrode sufficiently. That inception cloud gradually develops (image (*f*)) to the moment when it destabilizes into positive streamers (image (*g*)). The positive counter-streamers merge with the approaching negative streamers and develop into a highly ionized column (images (*k*) – (*n*)) with a diameter of about 40 cm. Two bright leaders (images (*m*) and (*n*)) grow towards each other from the electrodes through the ionized column, until they connect and form a spark. The radius of the fully developed negative corona exceeds 1 metre in Figure 2 (m). As we assumed earlier in [13], everything placed on similar distance influences the discharge development. In the present setup we moved such edges to a larger distance or covered them with conductive plastic, except for the grounded electrode.

*3.1.1 Inception processes*

The first faint light comes from the inception cloud around the HV electrode tip immediately after we apply the high-voltage. Positive and negative inception clouds have been found and investigated previously in [10–12]. Inception clouds of both polarities also appear in our experiments. They look similar although the current direction and electrode processes are quite different. Figure 3 (a) shows the negative inception cloud and the first streamers that emerge later from the disk rim. For this image we zoomed in on the HV electrode and enhanced the gain of the image intensifier. The camera shutter opening time is plotted in Figure 3 (b), superimposed on the voltage and current waveforms. The radius of the inception cloud $r_{ic}$ observed in Figure 3 (a) is 1.6 cm. From the breakdown electric field $E_{cr} \sim$ *32* kV/cm we estimate the necessary voltage: $U = r_{ic}E_{cr}$ and charge $Q = E_{cr} \times 4\pi\varepsilon_0 r_{ic}^2$ in the simple model of a spherical and conducting inception cloud. The resulting voltage $U = 50$ kV, the charge inside the cloud is $Q = 85$ nC. This is consistent with the measurements: At the beginning of the waveform the current $I_{HV}$ rises at the rate of 1 A/ns. Then the 85 nC charge is deposited in 13 ns or when the current attains 13 A. The measured voltage is indeed 50 kV at that moment. At the end of the camera exposure, the voltage has risen to 130 kV and the cloud has by then destabilized into streamers. The 120 A of observed current at the shutter closure time feeds the streamers emerging from the inception cloud and from the protection disk.

A positive inception cloud appears near the grounded electrode when the electric field at the grounded electrode tip is sufficiently large. Such an event happens at time t = 0.73 µs in Figure 3 (*b*). In this particular discharge, the cloud formation causes the sudden rise of the current $I_{GND}$ on the grounded electrode. Figure 4 shows two subsequent photos, each with 20 ns exposure, zoomed in on the grounded electrode tip, where both cameras have equal sensitivity. In the first photo, streamers emerge from the top of the positive inception cloud. The radius of the inception cloud is approximately 2 cm. The radius is consistent with the same $E_{cr}$ for the charge of 80 ± 50 nC derived from the current $I_{GND}$. The large spread is due to the noise in the measurement. Of course, the relation between charge and voltage cannot be maintained for the grounded electrode. The cloud is warped due to the field inhomogeneity caused by the tip and to the presence of the negative streamer front. In the second photo (*b*) the cloud is not visible anymore. Apparently the excitation processes are not dominant anymore. Only a bright spot of several mm size remains near the electrode tip. The electrode emits a continuous current into the streamer corona, as also mentioned elsewhere for such a process [10–12].

The streamers emerging from the negative inception cloud are relatively slow. With a speed of $4\cdot10^5$ m/s they cross a 13 cm distance from the cathode in Figure 3a. The voltage at this moment is about 140 kV, thus the streamers have a so-called stability field of 140/13 cm = 11 kV/cm in agreement with previous measurements [14]. For a recent critical discussion and new results on the theoretical understanding of the stability field, we refer to [15]. We varied the distance between the electrodes and found that the delay between negative and positive clouds depends on this distance. Figure 5 shows the delay for distances between 1.0 m and 1.5 m. Naturally the voltage at which the positive inception cloud appears increases with distance between the electrodes. Still we can interpret the increased delay figure as an averaged streamer propagation speed over the gap. The straight line is a linear fit with a slope of $4.4\cdot10^5$ m/s and a determination coefficient R-squared of 0.97. This averaged speed is remarkably close to the speed of the first streamers emerging from the inception cloud as just mentioned.

*3.1.2 Negative streamer propagation and streamer bursts*

We now analyze the negative streamer corona as shown in Figure 2 (*b*) – (*h*) as a half-sphere centered at the electrode tip. The radius of the corona sphere as a function of voltage is shown in Figure 6 (a), and as a function of time in Figure 6 (b). The straight line in panel (a) indicates so-called stability field $E_{min}$ = 12 kV/cm for negative streamers in STP air [16] and [15, figure 7]. Observations [14] show that the maximum streamer length is frequently determined by the applied voltage *U* divided by the stability field. But please note the experimental counter example in [17], with a theoretical substantiation in [15]. Figure 6 shows the second, third and fourth streamer burst, indicated by II, III, and IV. In each burst, the streamers grow up to the maximal length supported by the voltage at that instant. The applied voltage increases slower than necessary to support a continuous streamer development. As a result, the streamers stop momentarily as indicated by the current zeros mentioned in Fig. 2 and Sect. 3.1. But voltage keeps rising. The 50 ns delay between second and third streamer burst in Figure 6 (b) corresponds to a voltage rise of 100 kV. Streamers of the previous burst are then re-energized [9] but also new streamers emerge from the high-voltage electrode, in particular from the rim of the disk. The process repeats until the fourth and last burst bridges the gap. The radii can be fitted by a curve $r_c = 95 \cdot t^2$, where $r_c$ is the corona radius in cm and *t* is time in μs. The last five data points are above the stability field line and that bursts appear when the gap is being bridged.

*3.1.3 Branching angles of negative streamers.*

The images show that most negative streamers split into two branches during their propagation. An example is shown in Figure 7 (*a*). We analyzed how 500 branching events appear in the 2D image plane. We selected streamers with equal brightness on both branches without the gamma correction for the images. This selection emphasizes splits with both paths in a plane perpendicular to the camera viewing direction. Only outermost streamers were taken into account. The distribution of branching angles is shown in Figure 7 (*b*). The branching angle may depend on gas parameters and electric field. It is clear that in our case the negative streamer tends to split at an angle of 29º with 5º standard deviation. Interaction and branching angles of positive streamers in different electrical environments were analyzed in [17–19]. To our knowledge, branching angles of negative streamers were not reported before.

*3.2. Pre-breakdown in time-resolved image sequence*

To illustrate the importance of short exposure images, we compare in Figure 8 a long (a, 550 ns) and short (b, 25 ns) exposure. Many negative streamer traces are visible in Figure 8 (a), but most of them have no visible traces in (b) just after the end of (a). This means that most negative streamer channels may still exist as conducting channels, but lost their light emitting properties for instance by lack of growth and optically active streamer head. Only positive streamers rise from the inception cloud at the grounded electrode. The full discharge development is shown in Figure 9 in a time-resolved sequence. Panel (h) shows the currents and voltage taken at the discharge of panel (f). The single red-blue (RB) pictures (*a*) - (*g*) is an overlay of two images from two cameras, each of 50 ns exposure. They have been taken directly after each other. The image of the first camera is placed on the red layer of the picture, the second image on the blue layer. The light level is reproduced as color intensity, where a gamma correction of 0.2 has been applied for better visibility of lesser bright parts. The HV current jump at t = 0.4 μs already corresponds to the second streamer burst. From many other images not shown here, we find that the streamers bridge a distance of 30 cm with an average speed of $2 \cdot 10^6$ m/s, and then terminate at their maximal length allowed by the voltage, as discussed before.

Figure 9 (*a*) shows stages during the third burst accompanied by a current pulse with a maximum at t = 0.55 μs. The electrode distance is 127 cm as in Figure 2. In going from the corona outer extent to the electrode, one first observes that the outermost blue streamer traces are shorter than the red ones. These streamers heads become slower - from $1.7 \cdot 10^6$ to $1.1 \cdot 10^6$ m/s - and eventually stop moving as discussed before. The outermost streamer traces extend to a distance of 38 cm from the high-voltage electrode at the end of the second camera exposure. The corresponding current jump is less intense but lasts longer than at the second burst; this is a consistent behavior as it also shows up in Figure 2. The outermost traces appear as extension and branching of traces one layer deeper inside the corona, as has also been seen in Figure 2 again. The separation between outermost red and inner blue traces shows that the latter are re-ignited streamers of the previous burst. There is only little overlap between the red and blue traces. Apparently the light from the streamer heads stems mostly from short lived i.e. less than 10 ns exited atomic and molecular states. Finally, near the electrode the third burst appears in image (a). Here, the red and blue traces largely overlap. This suggests that these streamers propagate less, and new ones may emerge and old ones are re-exited by the increasing current necessary to feed the outer streamers. However, we cannot exclude a possible role of longer living excitations.

After a substantial rise of the voltage the fourth streamer burst initiates in the pre-ionized medium near the electrode at t = 0.7 μs (Figure 9 (*b*)). The outer streamers propagate in a higher electric field; their speed now varies between $2 \cdot 10^6$ and $4 \cdot 10^6$ m/s. Meanwhile, a barely visible positive inception cloud appears at the grounded electrode during exposure (b). In Figure 9 (*c*) the outermost streamers propagate farther in all directions over a half-sphere leaving diffuse and patchy structures behind. In part, these structures consist of positive streamers and ionized stationary traces that are reminiscent of "secondary streamers" [20,21]. The outermost negative streamers reach their final extent at an average distance between 80 and 100 cm from the cathode. The voltage then is about 1 MV, and the size of the corona is again consistent with the stability field (Sect. 3.1.2). Due to the proximity of the negative streamers, the electric field near the grounded electrode increases strongly and the inception cloud near the grounded electrode starts to emit positive streamers. In image (d) at t = 1 μs many positive upward moving streamers appear. The positive streamers transform into a structure that resembles a space leader [6] when they propagate through the zone pre-ionized by the previous negative streamer burst. The negative streamers that moved nearly horizontally from the high voltage electrode stop growing at this moment due to the rearrangement of the electric field by the positive charges from below.

Later, in image (*e*) the upward positive discharges merge with the ionized channels extending from the high-voltage electrode. Before (e) mostly the streamers heads emitted light. But now the channels become bright over their full length indicating strong Ohmic heating. These light emissions are also accompanied by a current pulse on the high-voltage electrode, often accompanied by high-frequency oscillations. These oscillations are similar to ones discussed earlier in [1] for positive discharges. The light emission concentrates in the cylindrical region confined between the electrodes. First it has an inhomogeneous structure (image e and f) where streamer traces are still visible. But later it becomes diffuse (image (*g*)) with only slightly perceivable conducting channels. All streamer traces outside this region vanish. Those streamers deposited negative charge around the gap, and the electrons have two options: flow back via the ionized channel or get attached or recombine with the ions. Within the light emitting and conducting region, two leaders grow towards each other from both electrodes. They concentrate the current in a narrower and more heated region. In Figure 9(*h*) the final breakdown and spark occur after time t = 2.3 µs.

*3.3 Detailed discharge structures near the high-voltage electrode*

The resolution of the later images in Figure 2 and 9 does not allow to resolve the details of the spaghetti-like discharge structures in the strongly pre-ionized region near the high-voltage electrode. When zooming in on that electrode, we observe several phenomena, a) light emitting fronts moving from and towards the electrode as well as b) stationary dots and c) stationary channels of varying brightness. Some of these have already been mentioned in [6]. In order to have consistent names we use 'streamers' of both polarities for a), and 'pre-leaders' for c) because these stationary structures may finally transform into a real leader. We interpret these pre-leaders as re-excited conducting remainders of negative streamers, also called "secondary streamers" in [20,21]. Items b) remain 'dots'. More theoretical and experimental investigations are required to understand the actual nature. The phenomena have been observed in many discharges and the images shown have been selected for clarity.

Figure 10 (*a*) shows positive and negative streamers moving towards and from the electrode; Figure 10 (b) gives the corresponding electrical parameters. Both cameras had 3 ns exposure time, and the blue image was delayed by 10 ns. The arrows indicate the displacement from the red to the blue images. A slight horizontal dislocation in Figure 10 (a) can be attributed to the different viewing angle of both cameras. The positive streamers move with a speed of about $2 \cdot 10^6$ m/s which is in a good agreement with the speed of positive streamers reported in [1]. These propagate towards the high-voltage electrode and towards the negative pre-leader which is attached to the electrode. Upon their encounter high-frequency electromagnetic oscillations are detected by the current probes. An example is shown in Figure 10 (*b*) at about t = 0.65 µs, marked as "collision". In [1] we observed that such encounters are associated with x-ray emission. The radial luminous intensity of all positive streamers observed here can be fitted to a Gaussian profile with FWHM of 2 ± 0.4 mm. A few discharge channels appear to be stationary, for instance both towards the lower left corner, one bright and one faint.

The positive streamers come in bundles. Figure 11 (a) again shows the area near the high-voltage electrode, now with longer exposures of 50 ns for each camera and a delay of 50 ns. The gain of the image intensifier has been reduced with respect to Figure 10 to avoid saturation. At least six tracks of positive streamer bundles can be seen. A stationary negative pre-leader extends from the right side of the high-voltage electrode. That pre-leader collects the right side bunch, and the electrode collects the left side bundle. The "collision" in Figure 11 (b) now indicates the moment the merging of positive and negative streamers, as in Figure 9 (e).

In the previous Sections we showed that the streamers propagate in steps, probably because of the relatively slow rise of the driving voltage. Many images zoomed-in on the high-voltage electrode show that the steps themselves behave irregular as well, at least the later stages of discharge formation such as the fourth steamers burst. Figure 12 gives several images taken at the time as in Figure 11 (b). The image intensifier gain is approximately equal to the one of Figure 10 again. Some tracks are attached to the high-voltage electrode; streamers appear as separated tracks as in Figure 9. The streamer luminosity varies over the track length. Some tracks split up into dots that often appear to be streamer branching points. The branching direction varies: most are downward, some upward. The branching direction depends on the streamer head polarity: positive ones moving upwards, negative downwards. We tentatively interpret the phenomena as follows: The luminosity variations over the track indicate irregular excitation and/or propagation. Charge accumulates in regions of more intense light, possibly even because of micro-stepping of the streamer. The accumulated charge or excitation promotes branching, similar to streamer emission from the inception cloud. The dots are then the visual stationary remainders of such accumulation.

The negative pre-leader does not form in a continuous way. The short exposure image with high image intensifier gain of Figure 13 shows variations of the pre-leader thickness. A half-millimeter thin channel with broad ends is visible in the middle of the image, indicated by the arrow. The high-voltage current rises gradually as shown in Figure 13 (*b*). A mix of streamers and their residues surrounds the composition. The camera's gain is set that high that leader and streamers are seen simultaneously, but the leader is overexposed into saturation. Since the saturation may lead to blooming in the CCD, the real thickness of the pre-leader may be smaller than observed.

*3.4 Leader phase*

Figures 14 (*a*) – (*c*) show three examples of the formation and development of bright structures that develop on top of the positive streamers from the ground electrode. The camera sensitivity and exposure moment are the same as in Figure 9 (*e*), but no gamma correction has been applied. The structure brightness is two up to four times larger than the streamer head traces. These structures are disconnected from the electrode, hence its name 'space leader' [6]. It is clear that the central part of the structure in Figure 14 (c) is brighter than its edges, indicating that the structure is stationary rather than moving. At both electrode the currents are almost equal to zero at this moment; compare with image (*e*) and (h) in Figure 9. So the electrodes transfer only a small amount of energy and charge into the gap region. The enhanced brightness should then stem from the electrostatic energy stored locally and from the charge redistribution and associated current density near and inside the space leader. The multiple channels of the space leader persist in the final spark, as shown by the sill photograph of Figure 14 (d).

Similarly bright leaders in Figure 2 (n) appear later inside the heavily ionized column. Figure 9 (g) shows that both leaders grow simultaneously and remain attached at both electrodes. These leaders are formed out of streamer channels that are heated by the increasing currents from both electrodes. Because of the gradual heating of an already existing channel, no 'leader velocity' can be defined here. The diameter of the column between two leaders in Figure 9 (g) is about 40 cm as mentioned in Section 3.1. In images without gamma correction, the optical activity appears concentrated to a shield of about 12 cm diameter around the leader. Figure 2 (o) shows the ultimate development into an over-exposed spark.

*3.5 Similarities between phenomena in the laboratory and in nature*

In natural lightning the negative leader proceeds by steps. Figure 15 demonstrates that stepped phenomena also exist in an air gap of 107 cm. We reduced the sensitivity of the cameras to see only the brightest parts of the discharge suppressing streamers and glowing channels. The negative pre-leader is attached to the high-voltage electrode in image (a). The structure under it extends to the leader channel and downwards in image (b). This is accompanied by a rapid change in the high-voltage current. The earlier current distortion at t = 0.88 µs is most likely associated with a similar process. An obvious similarity with natural lightning is the encounter of its downward leader and the upward leader originating at tall objects near the strike point. Our ground electrode acts as such an object.

Figure 16 compares a negative discharge in our laboratory (left) and a set of sprites observed by Cummer et al. [7] (right). The similarity in appearance and evolution is striking, in spite of the differences in size, environment and polarity. A sprite is one of the transient luminous events (TLE's) that can develop above a thundercloud when triggered by an intense lightning strike. Sprites extend between 40 to 90 km altitude where air density varies over 3 orders of magnitude. They also grow in both directions out of a bright 'space leader', where the ionosphere acts as diffuse electrode. Concerning the dots we quote [7]: "these dots are near branch points of the downward streamers, they also appear in locations where no branching of the downward streamers is visible". This is in line with our discussion on dots in Sect. 3.3. As far as can be judged form Fig. 16 (b), also the branching angle is quite similar to Fig. 16 (a).

## 4. Discussion and conclusions

In this study we described nanosecond-fast photography of the development of negative discharges in STP air combined with the determination of the electrical parameters. The voltage source is a standard lightning impulse generator connected to a spark gap with conical electrodes separated by 1 up to 1.5 metre. The applied voltage is 1 MV with a rise time of 1.2 µs. The discharge forms through a sequence of inception cloud, streamers, leaders and spark. The current measured on the high voltage electrode clearly shows up to four bursts of streamer formation at the HV electrode. Comparing the HV sources used in [6] and [22] the occurrence of the bursts can be readily attributed to our slower voltage rise time and higher source impedance which is primarily inductive. The time-integrated images show that next-burst streamers are a continuation of the earlier [9], and that new streamers are formed at the electrode. Outermost streamers of the first burst branch with the preferred angle of 29 ± 5 degrees. Their speed attains values between $2 \cdot 10^6$ and $4 \cdot 10^6$ m/s. This value is on order of magnitude larger than the average value derived from time-integrated images discussed in Sect. 3.1.1 and with Figure 5. The fourth burst is accompanied by increased current density and ionization near the HV electrode. Generally, the zoom-images near the electrode are chaotic. Streamer channels decay into 'dots', stationary items of increased luminosity. Such dots generate new streamer channels in both directions, away from and towards the HV electrode. Attached to the HV electrode luminous channels appear that transform into persistent pre-leaders with increasing brightness.

At the time of the fourth burst, the outermost streamers are halfway down the gap. The electric field at the grounded electrode has increased sufficiently to induce a positive inception cloud and streamers there. As for positive high voltage [1] these streamers proceed with less variation in current intensity. When the positive streamers merge with the negative corona, the current on the HV electrode decreases first. At the merging point the locally stored electric energy causes a 'space leader', a set of highly luminous and conductive channels that finally extend to the HV electrode. Then the current on both electrodes rises quickly and strong re-excitation of all channels takes place. The conductive column of 40 cm diameter forms and contracts to 12 cm when a few channels transform into hot leaders starting simultaneously at both electrodes. Full breakdown occurs when the leaders meet. Remarkably the space leader filaments persist even at half height in the breakdown phase.

In natural lightning leaders proceed by streamer emission [23]. It is worthwhile to investigate whether common factors with the laboratory experiments are 1) streamer and leader current restricted by the charge collection higher in the atmosphere and 2) micro-stepping of the leader through conversion of streamers into space leaders. As for the link with sprites we conclude that the visual resemblance with the chaos near the HV electrode and its growth pattern is striking, in spite of large difference in circumstances. The physical resemblance might be a matter of scaling [24].

**Acknowledgement:** This work was supported by the Dutch Technology Foundation STW under project BTP 10757.

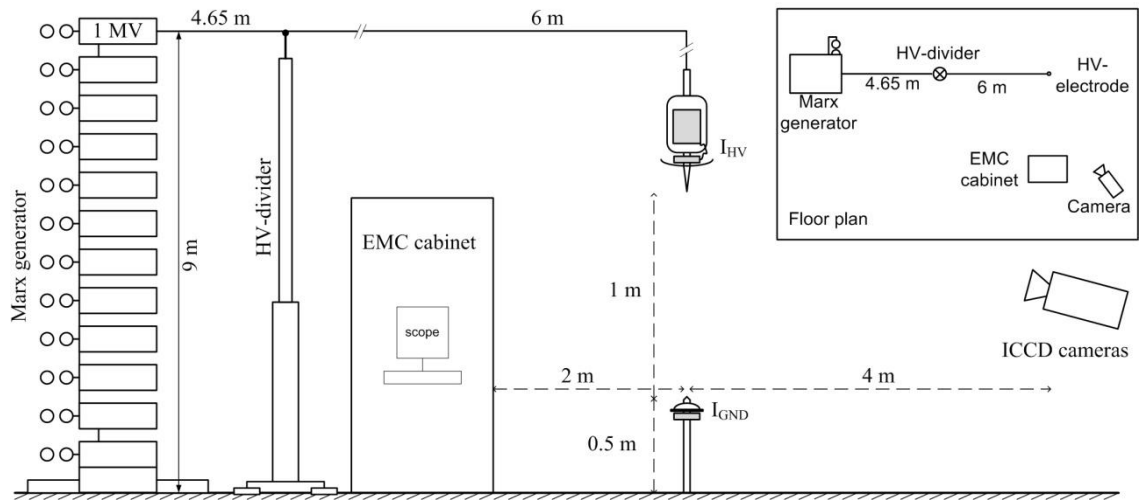

**Figure 1**. Schematic of the experimental setup. Two ICCD cameras are located at 3.5-4.5 m distance from the gap. The distance between Marx generator and the spark gap is 8 m. The upper right inset shows the scaled floor plan of the setup.

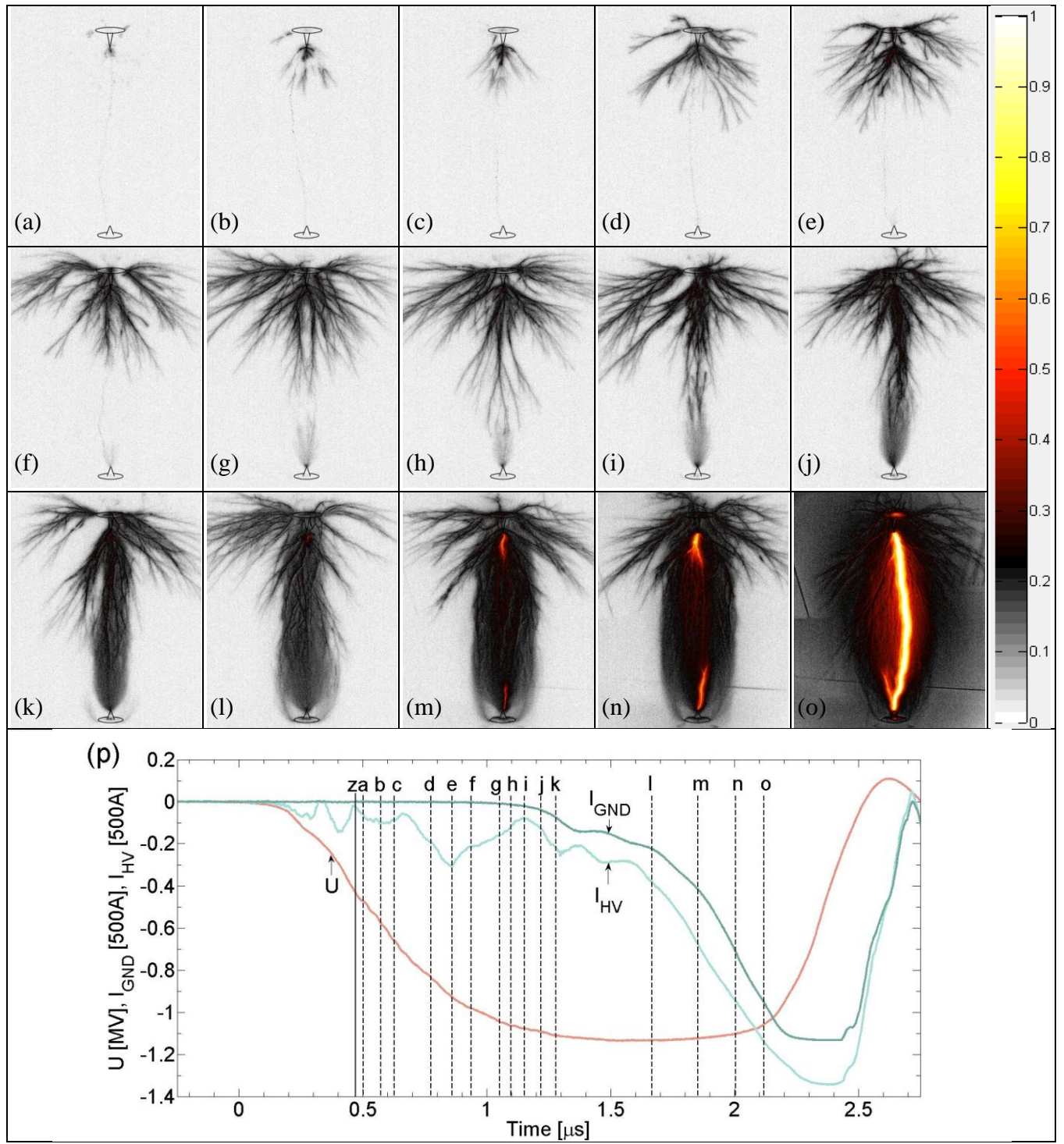

**Figure 2.** Time-integrated sequence of the development of a negative discharge over a gap length of 127 cm. Each picture shows a different discharge under the same conditions. The shutter always opens at $t = 0.47$ μs (solid line $z$). The exposure time varies from 25 ns in panel (*a*) to 1650 ns in panel (*o*). The linear color coding scheme for the light intensity is indicated on the right side of the figure. Voltage (U), cathode ($I_{HV}$) and anode ($I_{GND}$) current are represented in the bottom plot (p). These measurements are averaged over 65 discharges.

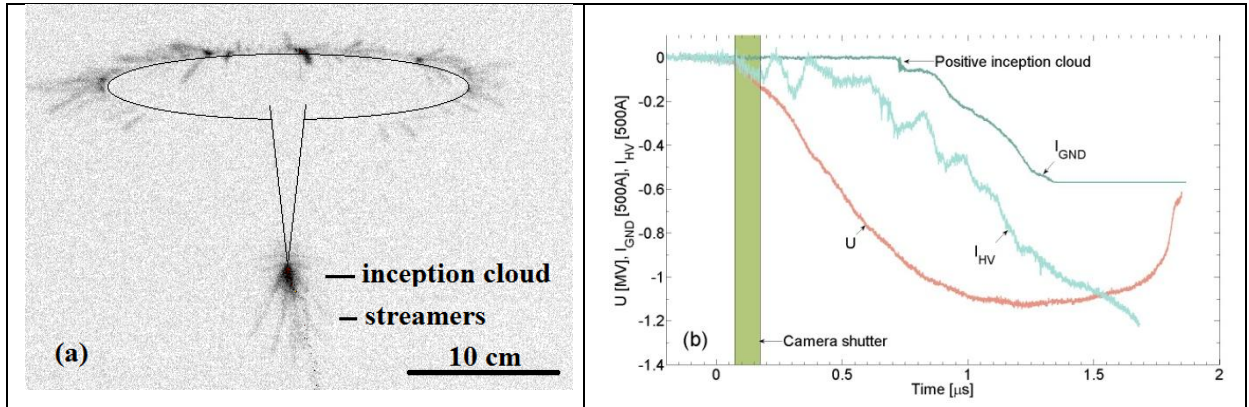

**Figure 3**. (a) Image of negative inception cloud and first streamer burst at the high-voltage electrode tip. The exposure time is 100 ns. The radius of the inception cloud around the electrode is 1.6 cm. Shutter opening time and moment of positive inception cloud formation are indicated in image (b). The gap distance is 107 cm.

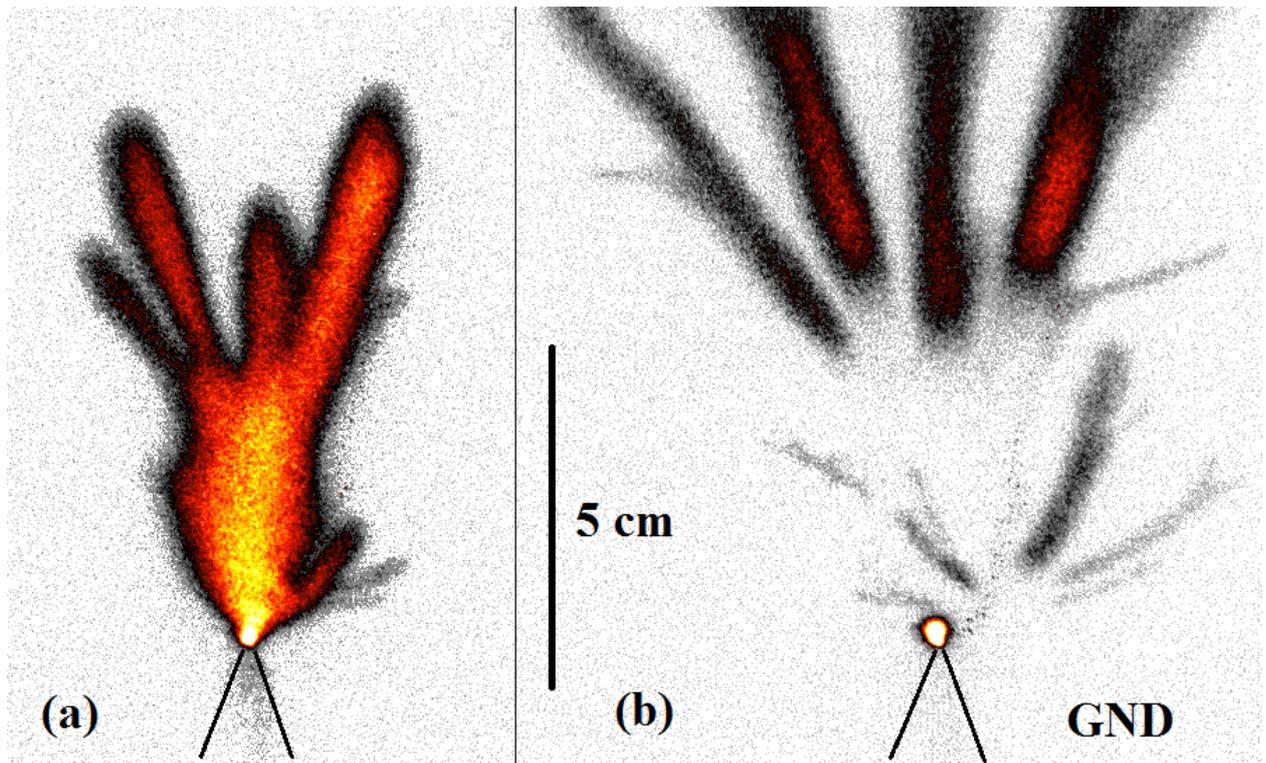

**Figure 4**. The positive inception cloud at the ground electrode is shown in two subsequent images, each with an exposure time of 20 ns. The inception cloud and the emerging streamers are visible in panel (a). Panel (b) shows the further growth of the streamers, and a bright dot of glow at the electrode that feeds the streamers and the now invisible cloud with current.

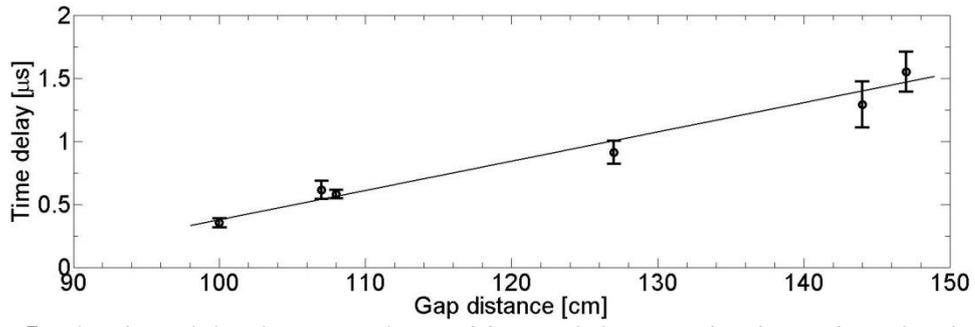

**Figure 5.** The time delay between the positive and the negative inception cloud as a function of the electrode distance, which can be interpreted as the averaged negative streamer propagation speed. The straight line with $4.4 \cdot 10^5$ m/s slope is a fit to the data.

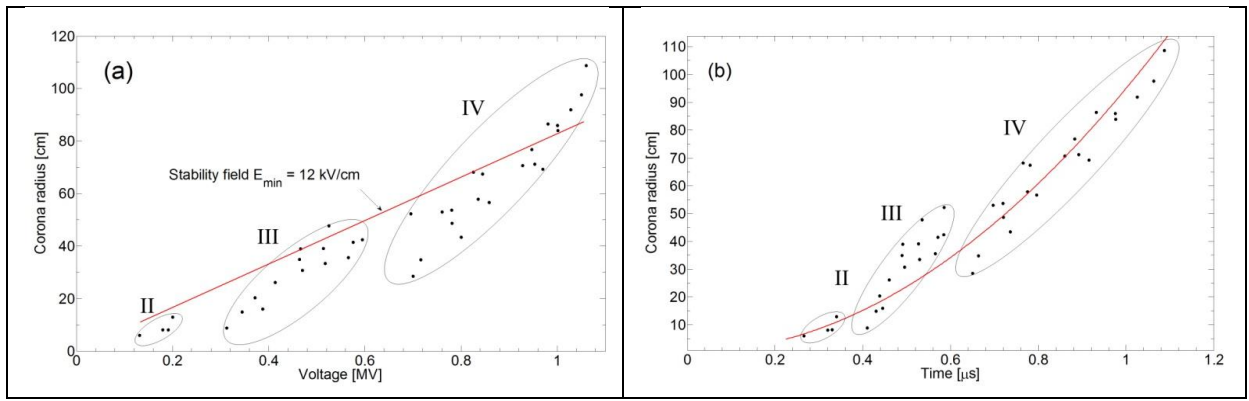

**Figure 6**. (a) The radius of the negative corona over the voltage in a gap of 127 cm length, obtained from 39 discharges. The so-called stability field of 12 kV/cm is indicated with a red line. The second, third and fourth streamer bursts are indicated with II, III, IV. (b) The corona radius now as a function of time. The time delay between bursts is 50 ns. A curve $r=95 \cdot t^2$ is fitted to the data, where $r$ is the corona radius in cm and $t$ is time in µs.

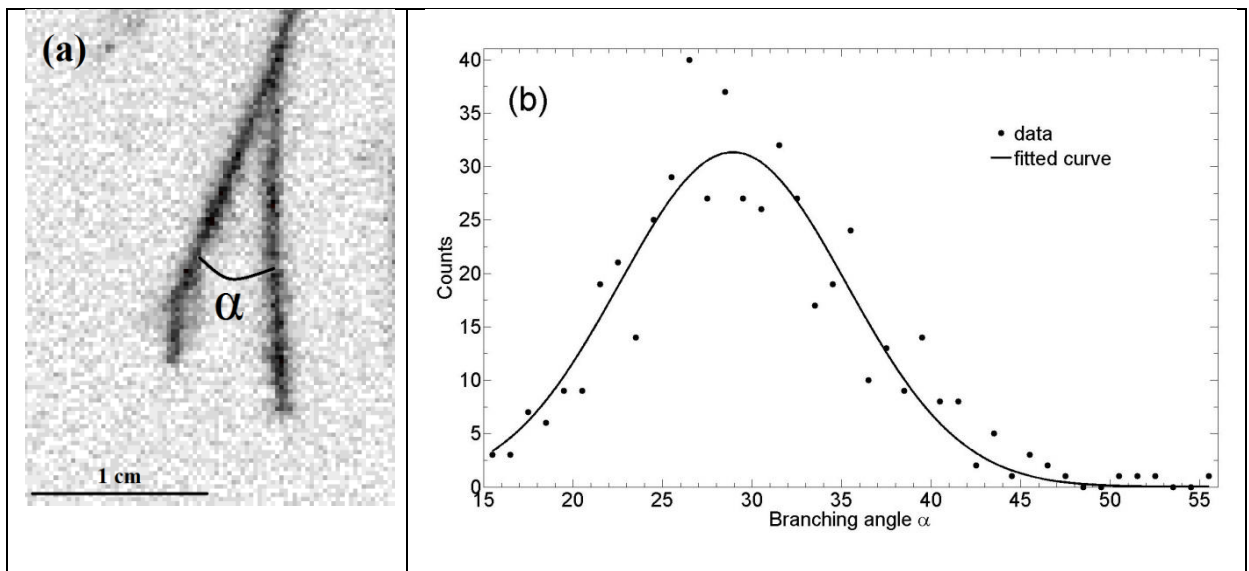

**Figure 7**. a) Negative streamer branching. b) The distribution of branching angles as it appears in the projection onto the image plane. The solid line indicates a Gaussian fit with mean angle $\alpha = 29º$ with 5º standard deviation.

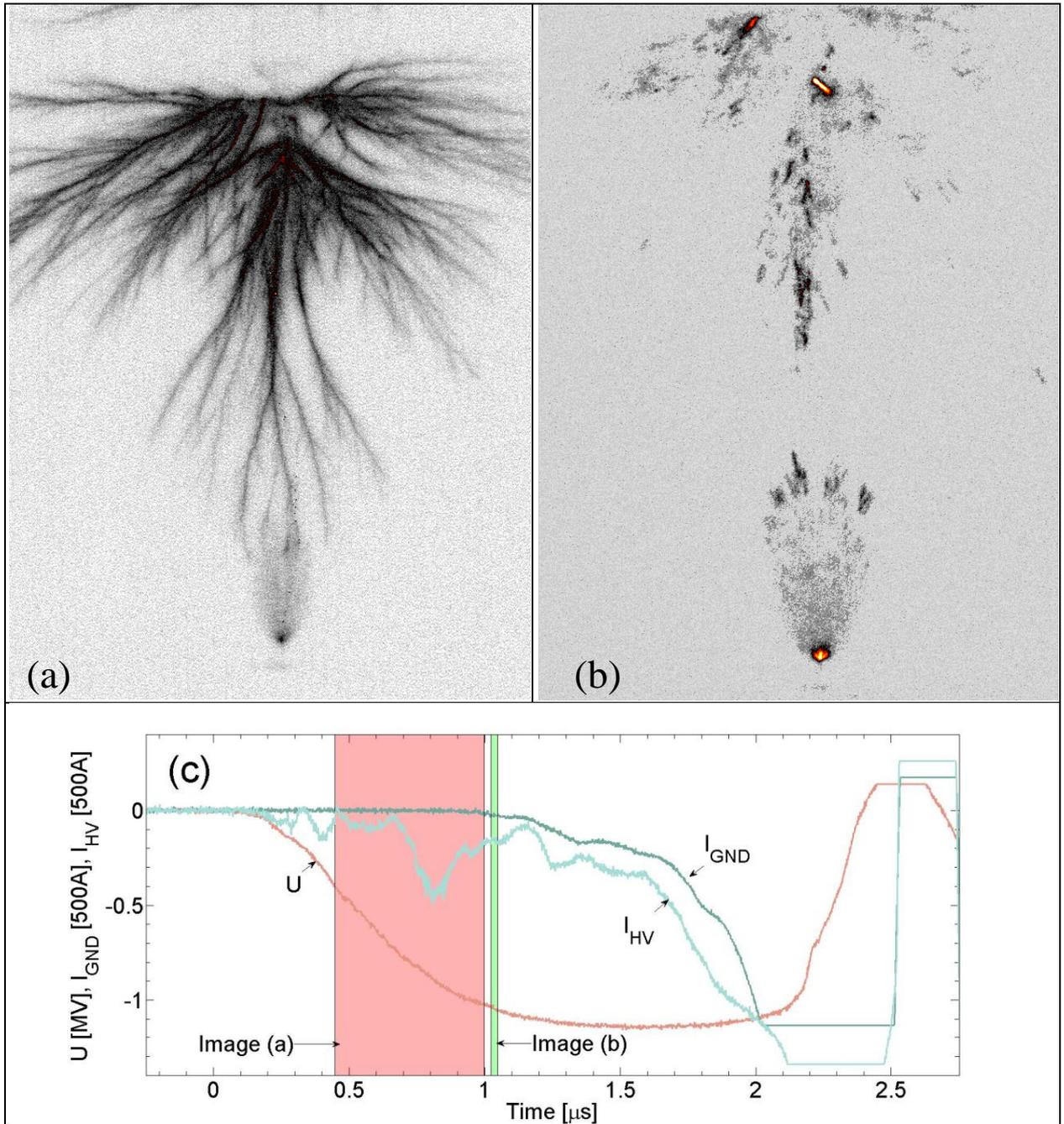

**Figure 8**. Long (a) and short (b) exposure images of pre-breakdown in a gap of 127 cm. Image (a) shows the light emitted during the fourth streamer burst between t = 0.65 μs and 1 μs. Image (b) shows the light from the ionization processes during a short time interval briefly later. The electrical characteristics of the discharge and the exposure times of the two images are indicated in panel (c). Because of the longer exposure time, image (a) is taken with a lower image intensifier gain than image (b).

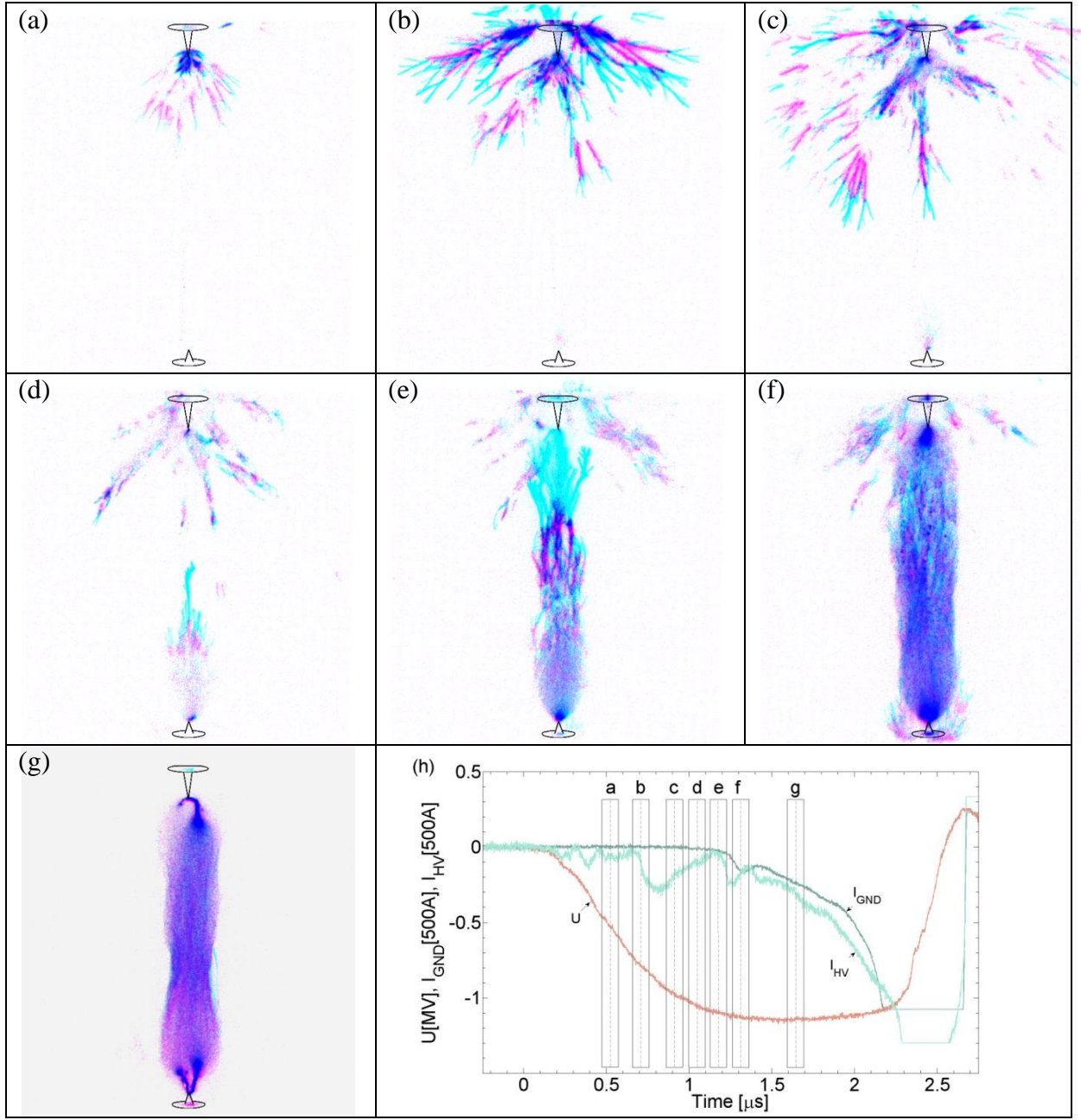

**Figure 9**. Time-resolved image sequence of the discharge development for an electrode distance of 127 m. Every red-blue picture from (*a*) to (*g*) consist of two consecutive images with 50 ns exposure time. The first image is placed on the red layer of the RB picture and the second one is placed on the blue layer. A gamma correction coefficient of 0.2 is applied to each picture. Panel *h* shows the electrode currents and voltage for panel (*f*).

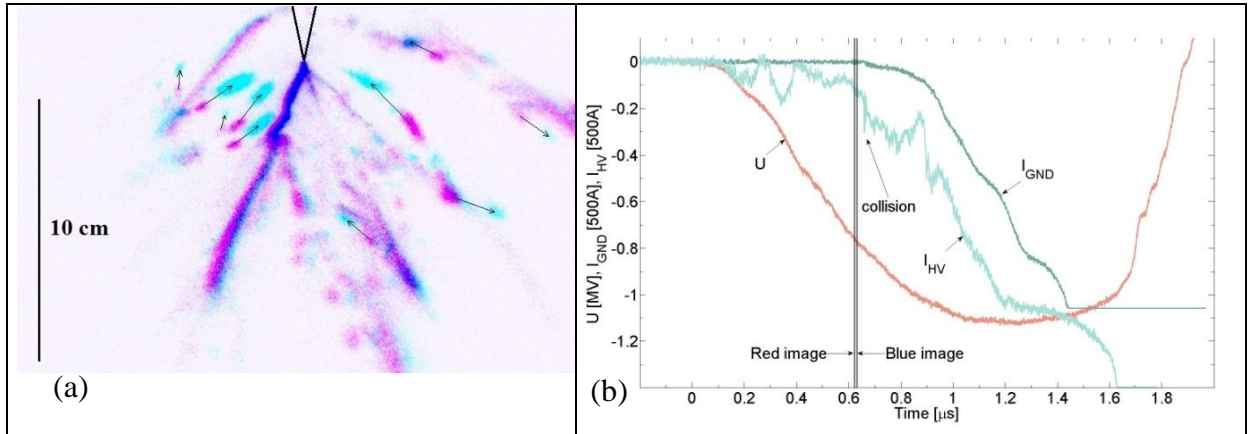

**Figure 10**. (a) Two images combined into one RB picture. The first image is placed on the red layer, and the second one on the blue layer. The exposure time is 3 ns for each image. The delay between the images is 10 ns. The gap distance is 107 cm. Arrows indicate the direction of motion, indicating that many "streamers" move towards the cathode. A bright stationary discharge channel (leader) is attached to the electrode tip. (b) Currents and voltage, where the moment of collision of the positive "streamers" of panel (a) with the cathode is indicated.

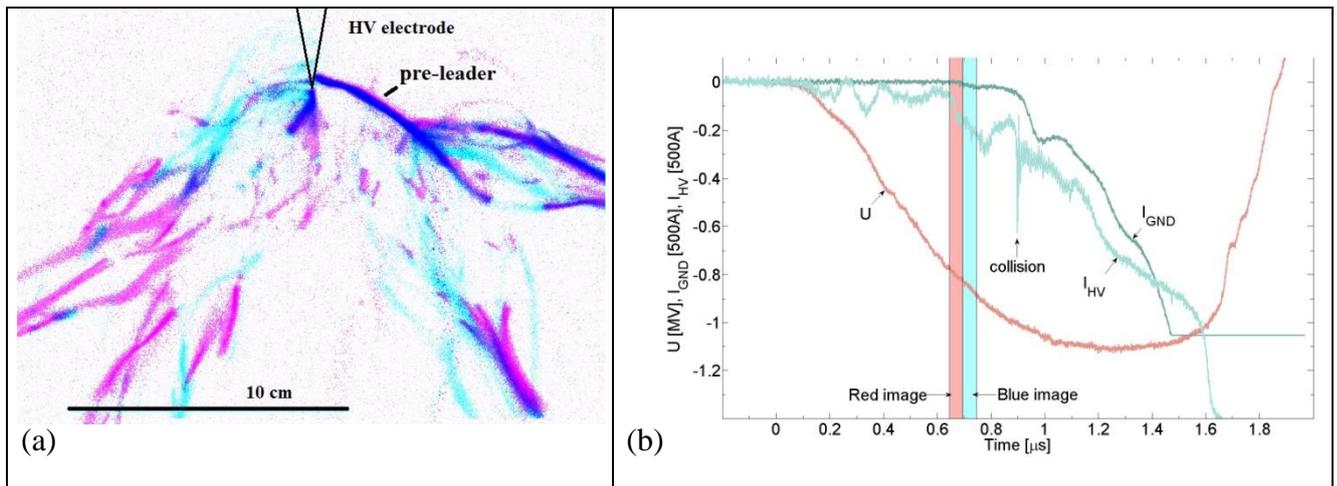

**Figure 11**. The RB picture consists of two images with 50 ns shutter opening time each. The first image is placed on the red layer of the RB picture and the second one is placed on the blue layer. A gamma correction coefficient of 0.2 is applied. The scales are the same as in Figure 10. The timing of the images is indicated in panel (b).

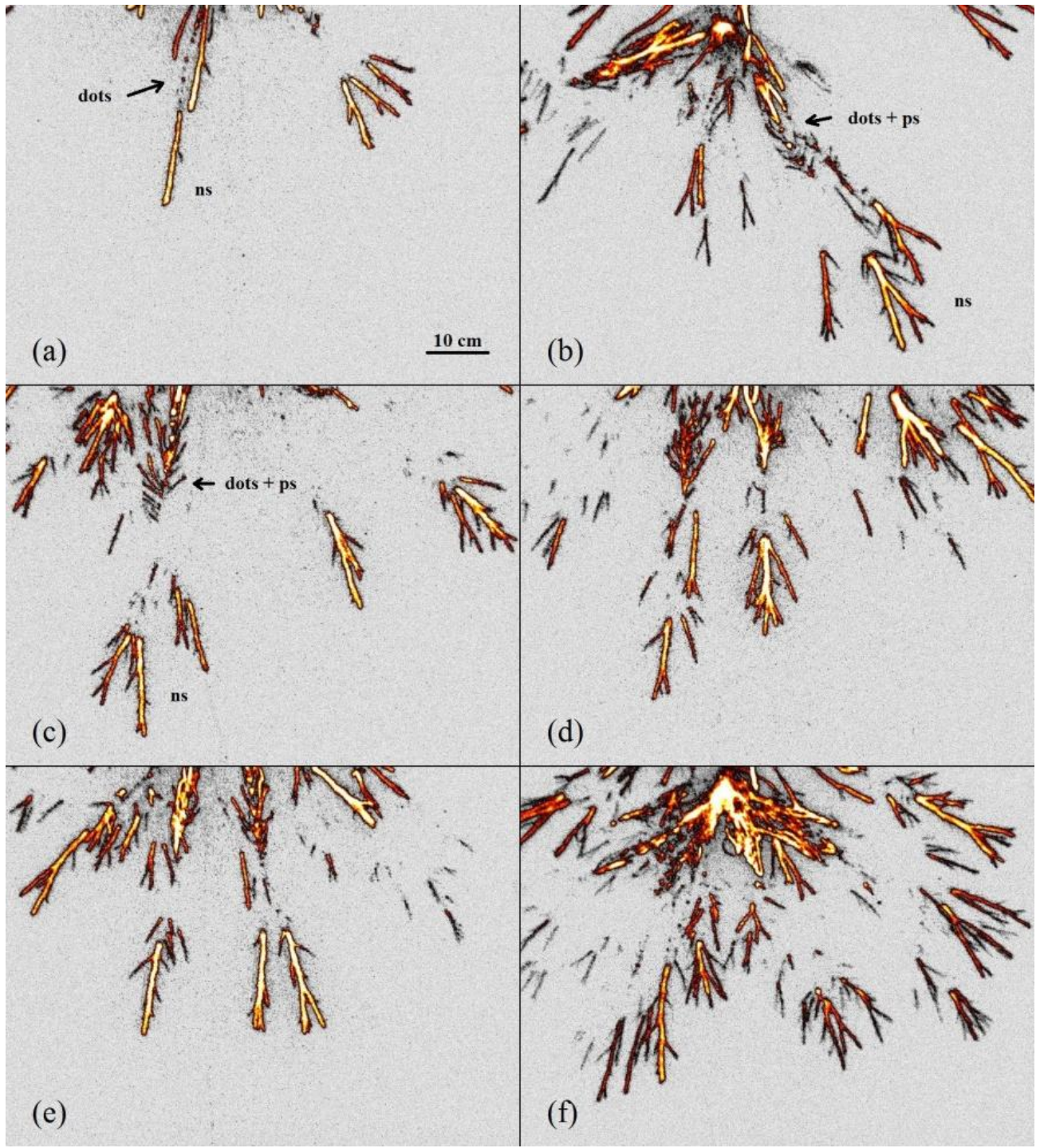

**Figure 12**. Typical images at the fourth streamer burst just below the high-voltage electrode. For all images time is the same as in Figure 11. Exposure is 50 ns. (a) Negative streamers (ns) leave isolated dots behind. The dots act as a starting point for new positive (ps) cathode-directed streamers (b) – (f). Image (a) clearly represents a less intensive streamer burst than image (f).

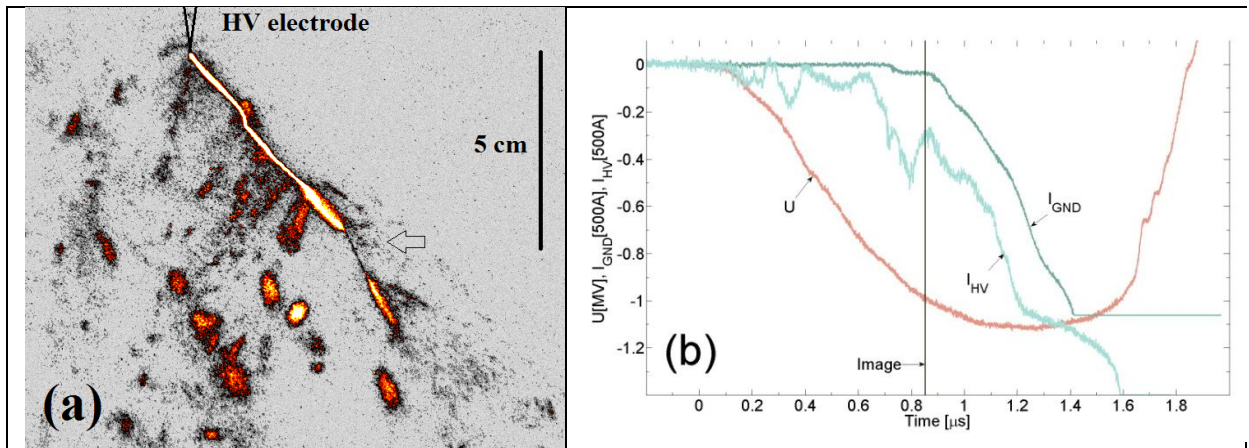

**Figure 13**. Fluctuations of the leader channel thickness. The gap distance is 107 cm, and the exposure time is 3 ns. The moment of the image is indicated in panel b that shows the electric characteristics of the event.

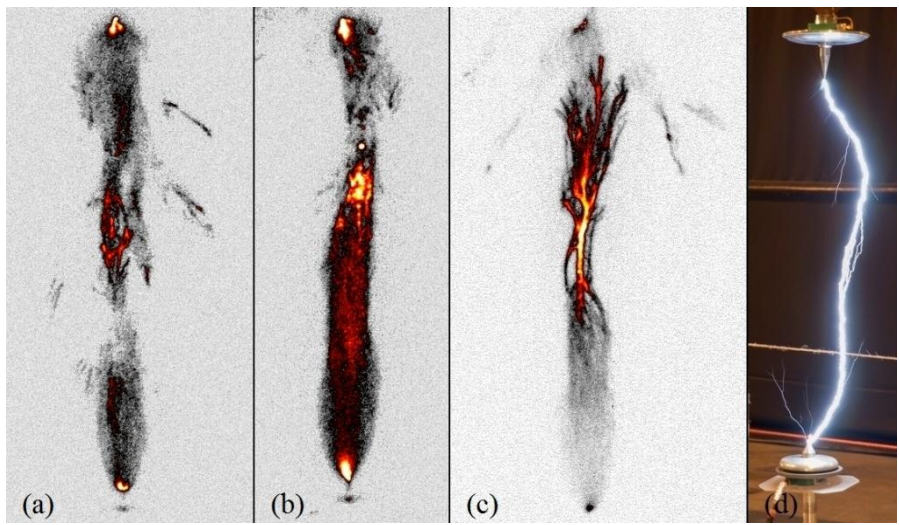

**Figure 14**. Space leaders observed in three different discharges. Images (a) – (c) were taken at the same moment as in Figure 9 (e). The exposure time is 100 ns, and the gap distance 150 cm. Image (d) is a photograph taken with shutter open during the whole discharge formation.

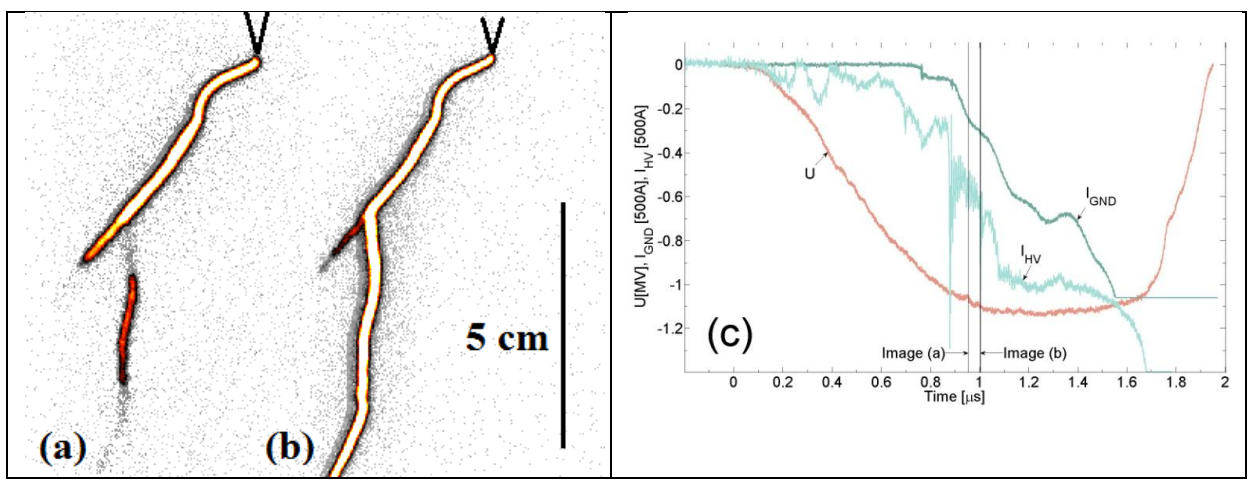

**Figure 15**. A stepping leader. The gap distance is 107 cm, and the exposure time is 0.5 ns for each image. The delay between the images is 50 ns. The electric characteristics in the right panel show that a peak in the high-voltage current is detected simultaneously with image (*b*) in the left panel.

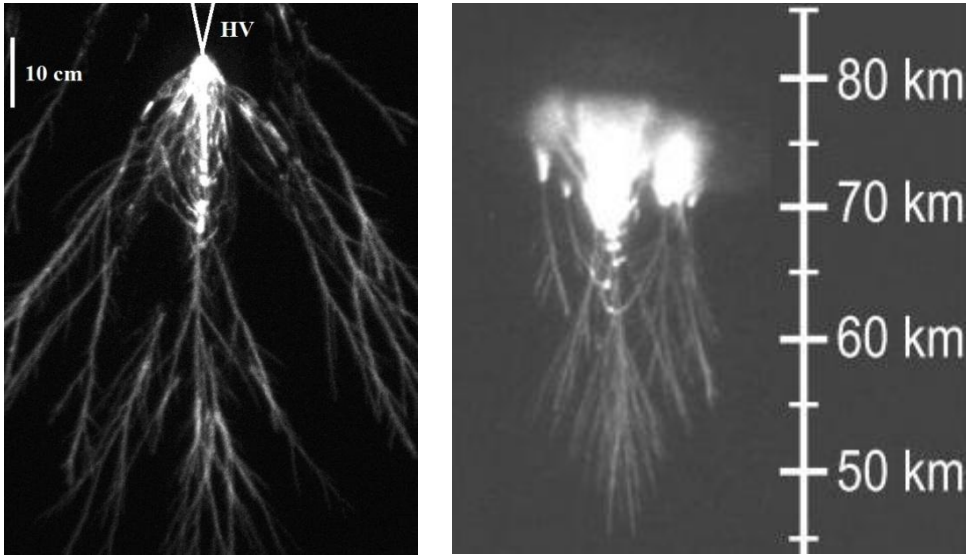

**Figure 16**. (a, left) Laboratory discharge with an exposure time of 100 ns, without color coding or gamma correction. (b, right) High speed image of a sprite discharge, adopted from [7].